# Astro2020 Science White Paper

# Debris Disk Composition: A Diagnostic Tool for Planet Formation and Migration

**Thematic Areas:**

☒ Planetary Systems  ☐ Formation and Evolution of Compact Objects

☒ Star and Planet Formation  ☐ Cosmology and Fundamental Physics

☐ Galaxy Evolution  ☐ Resolved Stellar Populations and their Environments

☐ Stars and Stellar Evolution  ☐ Multi-Messenger Astronomy and Astrophysics

**Principal Author:**
Name: Christine H. Chen
Institution: Space Telescope Science Institute (STScI)
Email: cchen@stsci.edu
Phone: (410) 338-5087

**Co-authors:**
Nicholas P. Ballering, Gaspard Duchêne, András Gáspár, Ludmilla Kolokolova, Carey Lisse, Johan Mazoyer, Amaya Moro-Martin, Bin Ren, Kate Y. L. Su, Mark Wyatt

**Co-signers:**
Elodie Choquet, John Debes, Julien Girard, Grant Kennedy, Quentin Kral, Meredith MacGregor, Brenda Matthews, Stefanie Milam, Marshall Perrin, Jason Wang

**Abstract**:
Debris disks are exoplanetary systems containing planets, minor bodies (such as asteroids and comets) and debris dust. Unseen planets are presumed to perturb the minor bodies into crossing orbits, generating small dust grains that are detected via remote sensing. Debris disks have been discovered around main sequence stars of a variety of ages (from 10 Myr to several Gyr) and stellar spectral types (from early A-type to M-type stars). As a result, they serve as excellent laboratories for understanding whether the architecture and the evolution of our Solar System is common or rare. This white paper addresses two outstanding questions in debris disk science: (1) Are debris disk minor bodies similar to asteroids and comets in our Solar System? (2) Do planets separate circumstellar material into distinct reservoirs and/or mix material during planet migration? We anticipate that *SOFIA*/HIRMES, *JWST*, and *WFIRST*/CGI will greatly improve our understanding of debris disk composition, enabling the astronomical community to answer these questions. However, we note that despite their observational power, these facilities will not provide large numbers of detections or detailed characterization of cold ices and silicates in the Trans Neptunian zone. *Origins Space Telescope* is needed to revolutionize our understanding of the bulk composition and mixing in exoplanetary systems.

# 1 Introduction

Debris disks, such as those around β Pictoris and HR 8799, are exoplanetary systems containing planets, minor bodies (such as asteroids and comets), and debris dust (see Figure 1, Lagrange et al. 2010, Marois et al. 2010). Unlike protoplanetary disks, debris disks contain very little natal, molecular gas (gas:dust ratios < 1), suggesting that the observed dust is continually replenished by collisions among minor bodies. Thousands of debris disk systems have been discovered to date, primarily from unresolved measurements of infrared excess (e.g. Eiroa et al. 2013, Su et al. 2006)- thermal emission larger than that expected from a bare photosphere at infrared wavelengths- using *Spitzer*, *Herschel*, and *Wide-field Infrared Survey Explorer (WISE)*. Follow-up, high-resolution, high-contrast imaging of the brightest systems has revealed warps (Heap et al. 2000), narrow rings (Schneider et al. 1999), and brightness asymmetries (e.g. Stapelfeldt et al. 2004). Since the gravitational influence of unseen planets can induce these structures (e.g. Lee & Chiang 2016, Wyatt et al. 1999), debris disks have been referred to as "signposts of planetary systems".

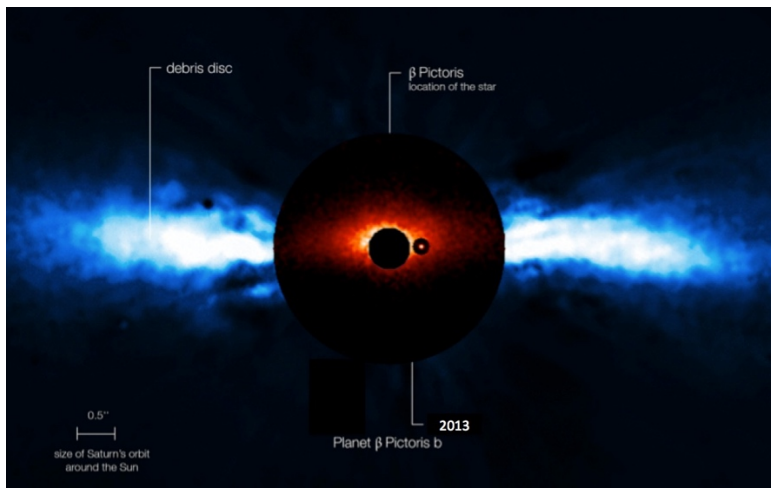

**Figure 1 Composite image of the β Pictoris planetary system showing the edge-on debris disk and the exoplanet β Pictoris b. The blue scale image shows the disk as observed by VLT/NaCo (Lagrange et al. 2010) and the red inset shows the Gemini Planet Imager (GPI) planet detection using the Spectroscopic mode (Macintosh et al. 2014) and the disk detection using the Polarimetric mode (Millar-Blanchaer et al. 2015).**

During the past twenty years, the detection and characterization of exoplanets has revolutionized our understanding of planetary systems (Winn & Fabrycky 2015). However, many questions remain about how planetary systems form and evolve and whether the history of our Solar System is common or rare. In planetary science, studies of minor bodies complement those of the terrestrial and giant planets and provide detailed constraints on how and when planets form and migrate (e.g. Kruijer et al. 2017, Johnson et al. 2016). Similarly, studies of debris disk dust properties have the potential to shed important light onto the processes by which minor bodies and therefore debris fragments form and evolve. For example, the detection of Obsidian and Tektite, glassy silicas altered at high pressures and temperatures, in the *Spitzer* Infrared Spectrograph (IRS) spectrum of β Pictoris Moving Group member HD 172555 led Lisse et al. (2009) to suggest that this system recently experienced a giant hypervelocity collision, such as the one that formed the Moon (see Figure 2).

The upcoming decade promises to be an exciting time for observational studies of debris disk dust composition with new thermal infrared (e.g. *SOFIA*/HIRMES, *JWST*) and visual scattered light facilities (e.g. *WFIRST*) coming on-line. Specifically, sensitive infrared spectroscopy will enable the first searches for solid state features in scattered light and thermal emission produced by ices and detailed characterizations of the composition, crystallinity, and grain size of detected grains. Space-based coronagraphs will continue to provide measurements of the visual scattered light color, enabling albedo measurements. However, despite our advances, many challenges still lie ahead. Specifically, the discovery and detailed characterization of solid-state thermal emission features from cold silicates and ices will remain out of reach for a statistically significant sample. Such measurements are needed to understand two basic questions about debris disks: (1) Are exoplanetary minor bodies really analogous to asteroids, Trans Neptunian Objects (TNOs), and comets in our Solar System? (2) Are there events that bring material from the outer planetary system into the terrestrial planet zone and vice versa?

## 2 Are Exoplanetary Minor Bodies Really Like Asteroids, TNOs, and Comets?

Little is currently known about the composition of dust in debris disks. Observations of the planets and minor bodies in our Solar System indicate that asteroids, formed within the snow line, are water-poor and compact while comets, formed outside the snow line, are water-rich and porous (Encrenaz 2008, A'Hearn 2011).

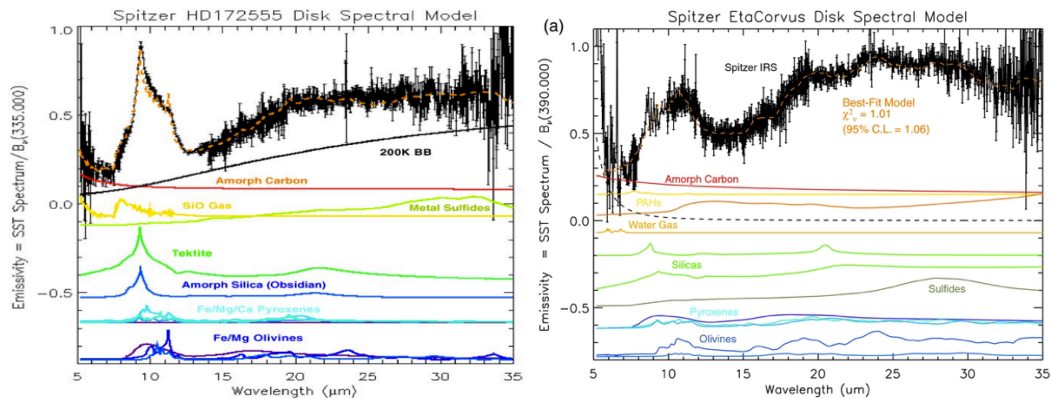

**Figure 2** *Spitzer* **IRS spectra of the HD 172555 (left) and η Crv (right) debris disks. The HD 172555 silicate emission feature shows evidence for glassy silicas (Obsidian and Tektite) and large quantities of small grains, indicative of a large, recent collision (Lisse et al. 2009). The η Crv silicate emission feature shows evidence for water- and carbon-rich dust in the terrestrial habitable zone, similar to that expected during a period of Late Heavy Bombardment (Lisse et al. 2012).**

The *Spitzer* IRS enabled mid-infrared spectroscopy (5 – 35 μm) of warm dust (with temperatures ~100 – 500 K) from large numbers of debris disks for the first time (Chen et al. 2014), including sensitive searches for solid-state silicate emission features at 10 and 20 μm. This unresolved spectroscopy discovered prominent silicate emission features toward approximately two dozen debris disks and weaker features around an additional 100 debris disks (Mittal et al. 2015). These observations indicated that (1) silicates are common at distances analogous to within Saturn's orbit and (2) the bulk of the infrared excess originates from grains that are too large and/or cold to produce silicate emission features at mid-infrared wavelengths. Although the Mid-Infrared Photometer for *Spitzer* (MIPS) and the *Herschel* Photodetector Array Camera and Spectrometer (PACS) provided SED and spectroscopic capability at far-infrared wavelengths (50 – 210 μm), these instruments were not sufficiently sensitive to search for solid state water ice at 63 μm and/or silicate emission at 69 μm for more than a handful of sources. Thus, the silicate and ice content of cold debris dust that forms the majority of the solid mass remains unknown despite the fact that this material is cold and therefore expected to contain large quantities of water ice. However, recently, CO gas and its photodissociation products (C II, O I, and CI) have been detected in ~20 debris disks using *Herschel* and ALMA (Matra et al. 2018).

To date, ~40 debris disks have been spatially resolved in scattered light (Choquet et al.

2018). Scattered light images break degeneracies in the interpretation of thermal emission SEDs by revealing the location of circumstellar dust. Scattered light images typically show dust at distances ~2 times further than estimated by fitting black bodies to thermal emission SEDs and assuming that the grains are large. In addition, scattered light fluxes can be compared with thermal emission fluxes to measure the dust albedo and infer composition without spectral features. For example, Rodigas et al. (2015) perform self-consistent modeling of the scattered light and thermal emission SEDs of the bright ring around HR 4796A using optical constants from 8 root materials (amorphous olivine, crystalline olivine, pyroxene, organics, tholins, water ice, iron, and troilite) and Mie Theory. They conclude that the disk is likely composed of silicates, organics, and iron. Solid-state features from these materials have not been detected. Indeed, the emissivities of organics and iron are constant as a function of wavelength and therefore these materials produce no solid-state emission features. The dust around HR 4796A is too cold to produce 10 and/or 20 μm silicate emission. Unfortunately, albedo is dependent not only on grain composition but also on grain size (Marshall et al. 2018). Therefore, inferring composition from SEDs alone is likely to remain frustrating.

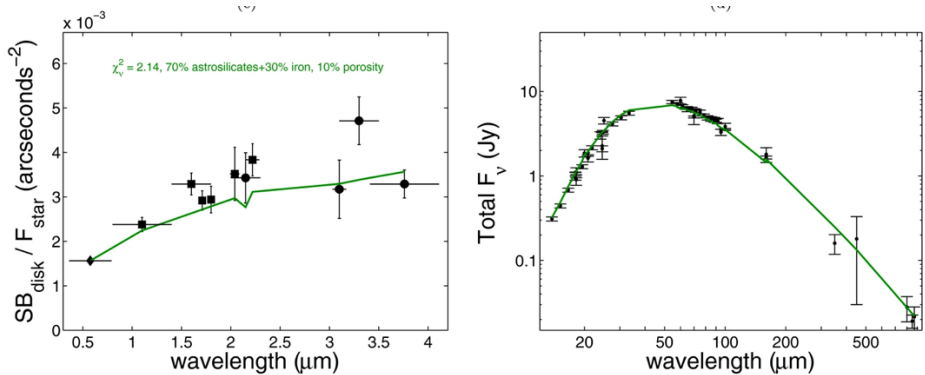

**Figure 3** HR 4796A scattered light (left) and thermal emission (right) SEDs. Model overlaid in green shows the best fit to both simultaneously using 70% silicates and 30% iron with 10% porosity (from Rodigas et al. 2015).

During the next decade, *SOFIA*/HIRMES and *JWST*/NIRSpec and MIRI and will enable scattered light and thermal emission, solid-state spectroscopy of ices and warm silicates; *JWST*/NIRCam and *WFIRST*/CGI will enable scattered light imaging at longer wavelengths and smaller inner working angles, respectively, than previously available. Specifically, HIRMES will provide unresolved, low-resolution spectroscopy (R~325-635) at 25-122 μm with sufficient sensitivity to search for water ice emission at 43 μm around a dozen bright debris disks ($F_\nu$(43 μm) > 0.5 Jy). NIRSpec and MIRI will enable sensitive infrared spectroscopy (R~100 – 3000) at 1 – 30 μm, including water ($H_2O$) and carbon dioxide ($CO_2$) frost features at 3 μm and silicate emission features at 10 and 20 μm. NIRCam will image disks at wavelengths up to 4.6 μm, enabling measurements of scattered light color at the longest wavelengths. CGI will provide contrasts of ~$10^{-9}$ at working angles between 3 – 20λ/D. Since its outer working angle is relatively small, CGI's imager (0.43-0.98 μm) and low resolution (R~70) Integral Field Spectrograph (IFS, 0.7-0.98 μm) will focus on dust in the inner regions of nearby planetary systems (see white paper by Mennesson et al.) and cold, Kuiper Belt dust for more distant objects (see white paper by Debes et al.) While these new facilities will enable additional searches for solid-state features and measurements of dust albedo, they will not provide a comprehensive understanding of the bulk composition of outer exoplanetary systems.

## 3 Radial Mixing in Exoplanetary Systems

In our Solar System, meteoritic evidence suggests that the formation of Jupiter may have divided the protoplanetary disk into two chemically distinct reservoirs (Kruijer et al. 2017) and that the migration of Jupiter through the asteroid belt may have mixed these reservoirs (Morbidelli et al. 2012). Specifically, isotopic studies of carbonaceous chondrites suggest that they accreted more water and refractory inclusions that other chondrites, consistent with their formation outside of Jupiter's orbit (Scott et al. 2018). These more primitive materials were mixed into the inner Solar System during the migration of Jupiter and Saturn as described in the Grand Tack Model. In the Grand Tack Model, Jupiter formed first and migrated inward in the disk through the asteroid belt, scattering asteroids outward. Once Saturn formed, Jupiter reversed it direction, migrating outward in the disk (locked in the 2:3 mean motion resonance with Saturn) through the asteroid belt, scattering asteroids inward until the gas dissipated (Walsh et al. 2011). Thus, the Grand Tack Model successfully explains the small mass of Mars and the Main Asteroid Belt, and the wide range of asteroid inclinations, eccentricities, and compositions of the asteroids.

Whether forming giant planets divide protoplanetary disks into chemically distinct reservoirs and whether the subsequent migration of these planets mixes these reservoirs in exoplanetary systems is not known. Some observations suggest that planet migration is occurring around other stars and mixing the reservoirs of small bodies. IRTF SpeX and *Spitzer* IRS observations of η Crv indicate the presence of ultra-primitive, water- and carbon-rich dust in the terrestrial planet zone of this 1.4±0.3 Gyr old F2V star. The impacting parent body in this system is presumed to be a TNO-like object because the *Spitzer* spectrum closely matches mid-infrared spectra of Ureilite meteorites from the 2008 Sudan Almahata Sitta fall; the parent body for which is believed to be a TNO. The present-day location of the dust is estimated from the inferred grain temperature. Detailed analysis of the *Spitzer* spectrum indicates that the impactor may have delivered as much as ~0.1% of the Earth's ocean to the habitable zone within this exoplanetary system (Lisse et al. 2012).

Spatially resolved thermal infrared spectroscopy and high contrast imaging total and polarimetric intensity images will allow observers to map dust properties as a function of stellocentric distance and therefore search for evidence of radial mixing in exoplanetary systems. For example, the *JWST* NIRSpec and MIRI possess Integral Field Spectrographs (IFSs) that reformat their fields-of-view into many slices that are subsequently spectrally dispersed. The NIRSpec Integral Field Unit (IFU) samples its 3″x3″ field using 0.1″ spaxels. The MIRI Medium Resolution Spectrograph (MRS) samples its 3″x3″-8″x8″ fields using 0.2″-0.6″ spaxels, depending on the channel used. However, recovering the spatial distribution of the dust from the IFS observations may be challenging because the host star is bright. Realizing the full potential of *JWST* will require the development of PSF subtraction techniques for instruments that are spatially and spectrally under sampled.

## 4 Into the Trans Neptunian Zone

Although *SOFIA*/HIRMES, *JWST*, and *WFIRST*/CGI will advance our knowledge of the composition of dust in debris disks, the discovery and characterization of cold silicates that are the remnants of planetary system formation in other planetary systems will still remain out of reach. For debris disks, thermal emission SEDS indicate that the bulk of the dust mass is located beyond the distance of Neptune where *SOFIA* lacks sufficient sensitivity and *JWST* has no diagnostic power. In our Solar System, the spatial demographics of TNOs has provided insight into the early dynamical evolution of our Solar System (e.g. Levison et al. 2008). Surveys, characterizing the surface colors of individual TNOs, may link objects to their formation locations (Pike & Lawler 2017). For exoplanetary systems, sensitive, high-resolution (R~4,000, Koike et al. 2003), far-infrared spectroscopy is needed to search for and characterize the shapes of the far-infrared, silicate emission features (40 – 70 μm). Higher resolution, far-infrared spectroscopy is needed to search for and characterize the dynamics of gas (see white paper by Matra et al.)

Forsterite has solid-state emission features at 33.8, 50, and 69.6 μm; enstatite has solid-state emission features at 40.7, 43.2, and 65.7 μm; and silica has a solid-state emission feature at 70 μm. Discovery of cold silicates via these emission features, in conjunction with measurements of water ice via scattered light at near-infrared wavelengths and/or thermal emission at far-infrared wavelengths (see white paper by Pontoppidan et al.), would provide constraints on the ice-to-rock ratio in exoplanetary system TNOs. Whether extrasolar TNOs should have ice-to-rock ratios as Solar System TNOs is not clear. *Spitzer* IRS spectroscopy indicates that while the terrestrial planet zones of T Tauri stars are water-rich those of Herbig Ae stars are dry (Pontoppidan et al. 2010). Far-ultra violet, stellar photons from early-type main sequence stars are expected photodesorb ices from the surfaces of micron to millimeter-sized grains out to distances of 150 AU (Gregorieva et al. 2007). Thus, the bulk of TNO-like, minor bodies in debris disks around early-type stars may be water-poor and silicate-rich.

Detailed characterization of the wavelengths and shapes of far-infrared silicate emission features can constrain polymorph composition, grain temperature, and the Mg/(Mg+Fe) ratio and in doing so provide evidence for radial mixing. For example, laboratory measurements of forsterite ($Mg_2SiO_4$) and fayalite ($Fe_2SiO_4$) indicate that the peak positions of the 30, 50, and 69 μm bands of forsterite shift toward longer wavelengths as the concentration of forsterite decreases (Koike et al. 2003). *Herschel* PACS moderate-resolution (R~1000) spectra of the β Pic debris disk revealed the presence of 69 μm forsterite emission, produced by magnesium-rich material (with $Mg_{2-2x}Fe_{2x}SiO_4$, x = 0.01±0.001) comprising 3.6±1.0% of the total dust mass, similar to that found in comets (de Vries et al. 2012). This material is more magnesium-rich than that found in asteroids (x=0.29). Asteroids possess more iron-rich silicates than comets because the activation energy of iron-rich silicates is higher and asteroids are heated to higher temperatures (Nakamura et al. 2011). Since β Pic's crystalline olivine was probably produced by thermal annealing, its presence in the outer regions of the disk indicates that the disk probably experienced some radial mixing.

## 5 Summary

During the upcoming decade, new facilities will enable measurements of debris disks solid-state features and albedos. These measurements will help to answer the questions: (1) Are debris disk parent bodies similar to asteroids and comets in our Solar System? (2) Do planets separate circumstellar material into distinct reservoirs? Unfortunately, we will continue to remain ignorant about the inventory of silicates in the Trans Neptunian region where the bulk of the solid mass resides because *SOFIA*/HIRMES lacks the sensitivity to detect the targets at sufficiently high signal-to-noise. Thus, a far-infrared mission, such as the *Origins Space Telescope* in the 2030s, capable of providing sensitive, moderate spectral resolution spectroscopy is needed to detect far-infrared, silicate emission features. The *Origins* Medium Resolution Survey Spectrometer (MRSS) has the resolution and sensitivity to search for and measure the line profiles of cold silicate emission features towards ~500 nearby debris disks discovered using *Spitzer* and *Herschel* and thus revolutionize our understanding of parent body properties and evolution.